\documentclass[11pt,twoside]{article} 
\usepackage{asp2004}
\usepackage{epsf}
\usepackage{psfig}
\usepackage{lscape} 

\begin{document} 
\title{A Search for Unresolved Double Degenerates Using IUE Archives}
\author{C.-P. Lajoie and P. Bergeron} 
\affil{D\'epartement de Physique, Universit\'e de Montr\'eal, C.P. 6128, Succ. Centre-Ville, Montr\'eal, Qu\'ebec, Canada, H3C 3J7}
\begin{abstract} 
We present preliminary results of a study aimed at detecting double
white dwarf systems using a method based on a comparison of optical
and UV spectra for 141 DA stars drawn from the IUE archives. In
particular, we are looking for dicrepancies between optical and UV
temperatures. Even though known unresolved degenerate binaries stand
out in this comparison, most temperature differences can probably be
attributed to the presence of reddening, or the presence of heavy
elements. We are in the process of securing additional optical
spectroscopic observations to increase the number of stars in our
analysis.

\end{abstract}


\section{Introduction}
Stellar evolution theory predicts that there should exist a
population of white dwarfs in binary systems with combined masses exceeding the
Chandrasekhar limit. If such systems have a short enough orbital period,
merger within a Hubble time will make them possible candidates for
type Ia supernovae progenitors.  The problem is that very few double
degenerates (DDs) are actually known to account for the observed type
Ia supernovae rate. We present the results of a possible new method for
detecting DDs based on the comparison of optical and UV spectroscopic
temperatures. In particular, we are looking for discrepancies between
these two temperature estimates, which may indicate the presence of
more than one white dwarf in the system.

\section{Sample and Mass Distribution}
Our sample consists of DA stars drawn from the IUE archive data of
Holberg et al. (2003).  So far we have secured optical
spectra for 141 objects, and we plan to secure similar data for all white dwarfs in the Northern hemisphere. The effective
temperatures and surface gravities are first obtained by fitting
simultaneously the H$\beta$ to H8 Balmer line profiles using the
spectroscopic technique described in Bergeron et  al. (1992). 
Our model atmospheres assume a pure hydrogen composition, take into account NLTE 
effects and include convective energy transport. Masses are also obtained using 
the evolutionary models of Wood (1995) with thick hydrogen layers. Figure 1 presents 
the mass distribution as a function of $\log T_{\rm eff}$. Surprisingly, our sample contains 
very few low-mass stars, in contrast with the results of other spectroscopic
investigations.  Our sample contains only two very low mass objects, WD 1022$+$050 
(LP 550$-$52) a known double degenerate (Maxted \& Marsh 1999), and WD 0943$+$441 (G116$-$52).

\begin{figure}[!ht]
\plotfiddle{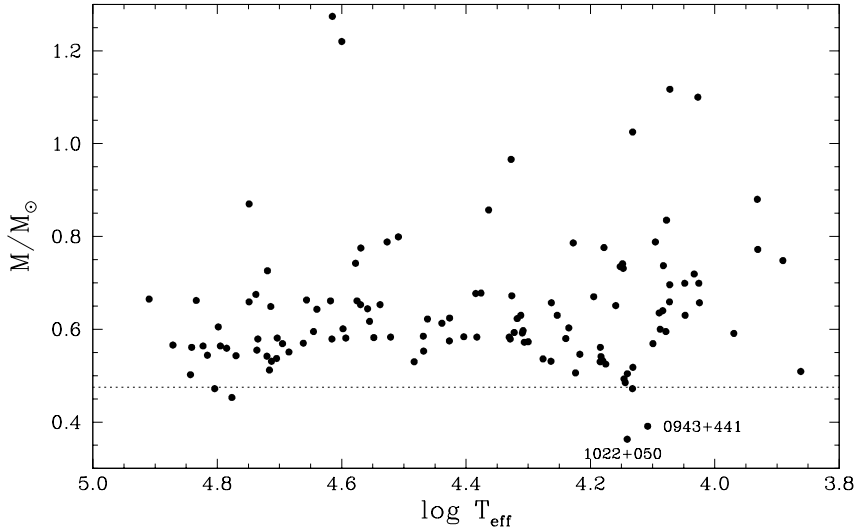}{6.3cm}{-90.}{50.}{50.}{-200.}{255.}
\caption{Distribution of spectroscopic masses as a function of effective temperature for
our complete sample.  The stars below the dotted line are most likely binary systems since
single white dwarfs with masses lower than $\sim$0.47 M$_\odot$ cannot have been procuced within 
the lifetime of the Galaxy.  They should also have a helium core.} 
\end{figure}

\section{UV Effective Temperature Determinations}
Since the UV energy distribution is mostly independent of $\log g$, we
constrain the $\log g$ value from the optical solution, and then use the
UV spectrum to derive a temperature from two different methods. The
first one relies on the slope of the energy distribution; here only
relative fluxes are important and the effective temperature and the
solid angle are considered free parameters. The second method, based
on the approach of Finley et al. (1990), relies 
on the $V$ magnitude.  Here, the model fluxes are normalized at $V$, and 
the measurement thus relies on the accuracy of the absolute UV fluxes 
and $V$ magnitudes. Depending on each star, one method could be preferable 
to the other, since there are instances when the IUE absolute fluxes are not reliable, 
or when the $V$ magnitude value is uncertain or suffers from contamination 
by a companion.

\section{Results}

The results of both methods are compared in Figure 2 as a function of distances,
which are obtained from the optical value of $\log g$ ---from which we derive
$R$--- combined with the UV solution for the solid angle $\pi(R/D)^{2}$.  As can
be seen, the two methods produce the same overall results.  The dotted lines represent the $\pm 10\%$ difference between optical and UV temperatures,
which we consider as a conservative estimate of the uncertainties of the methods. 

Our results first indicate that UV temperatures are
generally underestimated with respect to optical temperatures for hot
stars ($T_{\rm eff}$$>$40,000 K) and for distant stars ($D>100$ pc; not shown here). This
suggests that the presence of heavy elements in the atmosphere of hot
stars, or the effect of interstellar reddening, or both, may play an
important role in explaining our results. For the moment, however, we
simply restrict our analysis to nearby objects ($D<100$ pc) for which
reddening is negligible.  The
optical and UV temperatures all agree within $10\%$, with the exception
of the white dwarfs discussed in the next paragraph.  

\begin{figure}[!ht]
\plotfiddle{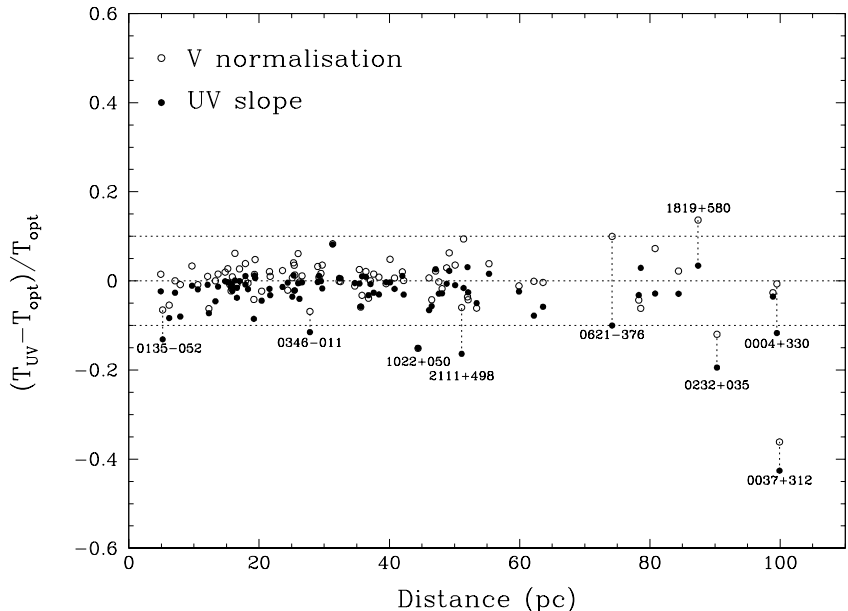}{7.cm}{-90.}{48.}{48.}{-185.}{248.}
\caption{Differences between UV and optical temperatures for stars located within $100$ pc using both methods for fitting the UV energy distributions.  Within such a limit, reddening is assumed to be negligible.}
\end{figure}

\par Two DD systems are identified in this plot, WD 0135$-$052 (L870$-$2, EG 11)
and WD 1022$+$050 (LP 550$-$52) discussed above and also shown in Figure
1. These are the only two known DD systems in this plot, both of which 
show discrepant optical and UV temperatures, indicating that our method 
works. The fact that only two such systems are identified here may 
not be surprising since DDs have been mostly discovered by surveying 
low mass DA stars, only two of which are found in our sample.  

\par The remaining objects outside our $10\%$ cutoff all have optical
temperatures in excess of 40,000 K: WD 0346$-$011 (GD 50), WD 2111$+$498 (GD
394), and WD 0004$+$330 (GD 2). All have consistent $T_{\rm eff}$ values when the $V$
normalization method is used, but discrepant values when the UV slope
method is used. For WD 1819$+$580, the reverse is true.  
WD 0232$+$035 (Feige 24) has an M dwarf 
companion and all three temperature estimates may be uncertain. The last two
objects, WD 0037$+$312 (GD 8) and WD 0621$-$376 exhibit quite different optical and UV temperatures,
with both methods. These objects clearly deserve further investigation.

\section{Simulations}
Finally, it is instructive to simulate what results could be expected
in the presence of DDs with arbitrary effective temperatures and
surface gravities. To do so, we first begin by co-adding two synthetic
spectra with various values of $T_{\rm eff}$ and $\log g$, properly weighted by
their respective radii. The results are then analyzed with the optical
spectroscopic and UV slope methods as if they were single stars.  The results
of our simulations are displayed in Figure 3 together with the results for our
sample of 141 DA stars.  It is important to note that the 
best fits do not necessarily represent good fits.  For instance, the points above and below $40\%$ would have been ruled out based on the quality of the solution (since both cool and hot optical solutions are considered here, below and above the maximum equivalent width  near 13,000 K).  Our 
results indicate that (1) double degenerate systems such as those simulated
here do not exist in large numbers and (2) that the discrepant
temperatures for the hot stars shown in Figure 2 cannot be explained
in terms of binarity.

\begin{figure}[!ht]
\plotfiddle{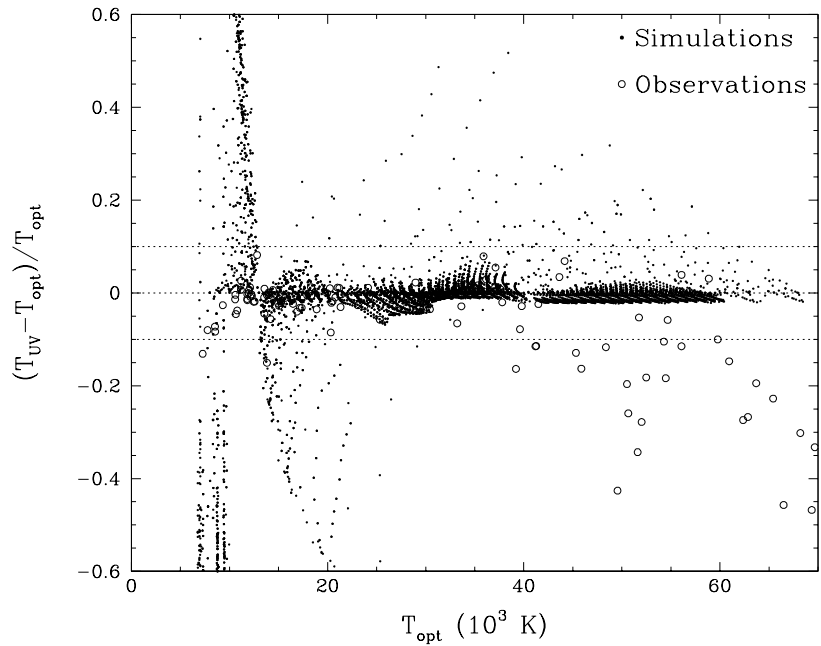}{7.8cm}{-90.}{48.}{48.}{-190.}{255.}
\caption{Results of our simulations.  Small dots are the simulated temperature
differences of two coadded and fitted synthetic spectra with arbitrary 
effective temperature and surface gravity, while open circles are the estimated
temperatures differences for our complete sample of stars.}
\end{figure}

\acknowledgements{This work was supported in part by the NSERC Canada 
and by the Fund FQRNT (Qu\'ebec).}

\end{document}